\pdfoutput=1

\documentclass[fleqn,10pt]{wlscirep}
\usepackage[utf8]{inputenc}
\usepackage{amssymb}
\usepackage[T1]{fontenc}
\usepackage{bm}
\usepackage{nameref}
\title{The germanium quantum information route}

\author[1,*]{Giordano Scappucci}
\author[2]{Christoph Kloeffel}
\author[3]{Floris A. Zwanenburg}
\author[2]{Daniel Loss}
\author[4]{Maksym Myronov}
\author[5]{Jian-Jun Zhang}
\author[6]{Silvano De Franceschi}
\author[7]{Georgios Katsaros}
\author[1]{Menno Veldhorst}
\affil[1]{QuTech and Kavli Institute of Nanoscience, Delft University of Technology, PO Box 5046, 2600 GA Delft, The Netherlands}
\affil[2]{Department of Physics, University of Basel, Klingelbergstrasse 82, CH-4056 Basel, Switzerland}
\affil[3]{MESA+ Institute for Nanotechnology, University of Twente, P.O. Box 217, 7500 AE Enschede, The Netherlands}
\affil[4]{Department of Physics, University of Warwick, Coventry CV4 7AL, United Kingdom}
\affil[5]{National Laboratory for Condensed Matter Physics and Institute of Physics, Chinese Academy of Sciences, 100190, Beijing, China}
\affil[6]{Univ. Grenoble Alpes and CEA, IRIG/PHELIQS, F-38000 Grenoble, France, }
\affil[7]{Institute of Science and Technology Austria, Am Campus 1, 3400, Klosterneuburg}

\affil[*]{e-mail: g.scappucci@tudelft.nl}

\begin{abstract}
In the worldwide endeavor for disruptive quantum technologies, germanium is emerging as a versatile material to realize devices capable of encoding, processing, or transmitting quantum information. These devices leverage special properties of the germanium valence-band states, commonly known as “holes”, such as their inherently strong spin-orbit coupling and the ability to host superconducting pairing correlations. In this Review, we initially introduce the physics of holes in low-dimensional germanium structures with key insights from a theoretical perspective. We then examine the material science progress underpinning germanium-based planar heterostructures and nanowires. We review the most significant experimental results demonstrating key building blocks for quantum technology, such as an electrically driven universal quantum gate set with spin qubits in quantum dots and superconductor-semiconductor devices for hybrid quantum systems. We conclude by identifying the most promising prospects toward scalable quantum information processing.

\end{abstract}

\begin{document}

\flushbottom
\maketitle

\thispagestyle{empty}
\section*{Introduction} \label{sec:Introduction}

The first quantum revolution has provided a microscopic understanding of nature underpinning the development of transistors. These were first made in germanium (Ge) due to its appealing electrical properties. Silicon (Si) took over as the material of choice for microelectronics because of the exceptional quality of silicon dioxide, allowing for the fabrication and integration of increasingly smaller transistors. Two key breakthroughs have fuelled a Renaissance in Ge-based materials and technologies\cite{Pillarisetty2011academic,KAMATA2008}. First, the maturity achieved by Ge-compatible high-$\kappa$ dielectrics overcomes the lack of a naturally stable dielectric; secondly, the heterogeneous integration of Ge on Si within a conventional CMOS process bypasses the need of  developing Ge substrates at prohibitive manufacturing costs. Germanium has the highest hole mobility of semiconductors at room temperature and therefore is considered a key material in a \textit{More than Moore} approach to extend chip performance in classical computers beyond the limits imposed by miniaturization.

Because of the unique combination of intrinsic material properties and compatibility with existing CMOS technology, Ge is also emerging as a promising material within the current second quantum revolution, in which quantum matter is engineered to develop disruptive technologies beyond the reach of classical understanding.  Quantum confined holes in Ge are a compelling platform for quantum technology based on spins\cite{Loss1998}, topological states\cite{Kitaev_2001}, and gate-controlled superconducting qubits (gatemons)\cite{larsen_semiconductor-nanowire-based_2015}.  

The high hole mobilities benchmark Ge as an ultra-clean material platform for well-controlled, high-quality quantum dots. The low effective mass, tunable by confinement and strain, gives quantum dots with large energy level spacing allowing to relax lithographic fabrication requirements. Uniformity and ease of fabrication are critical for scaling up to large quantum systems. Long hole-spin lifetimes are expected for two reasons: first the hyperfine interaction is suppressed due to the $p$-type character of the valence band; secondly, Ge can be engineered by isotopic purification into a nuclear spin-free material. Other important properties of holes in Ge are the large and tunable $g$-factors and spin-orbit interaction energies. These are key ingredients for fast all-electrical spin qubit control without the need of any microscopic structure, for spin qubit coupling at a distance via superconductors, and for the emergence of Majorana zero modes (MZM) for topological quantum computing. 

From a fabrication perspective, virtually every metal on Ge shows a Fermi level pinned close to the valence band\cite{Dimoulas2006}, including superconductors. As a consequence, it is straightforward to make ohmic contact to confined holes in Ge, without the need for local doping or implantation with associated high thermal budget. Furthermore, the low Schottky barrier at the metal/semiconductor interface facilitates the formation of transparent contacts to superconductors, a key building block in semiconductor-superconductor hybrids. Most importantly, Ge is a foundry-compatible material enabling advanced device manufacturing and integration. This is crucial for advancing to large-scale quantum systems as many challenges related to epitaxy, dielectrics, and variations of critical device dimensions may be solved by resorting to the advanced process control in a state-of the art manufacturing facility\cite{pillarisetty_qubit_2018}.

These attractive prospects have motivated extensive research efforts in the past years. Advances in Ge-based materials and physical understanding have led to  impressive achievements that we address in this Review. We shall specifically focus on hole-based devices with relevance to the developing field of quantum computation. Three materials platforms have emerged as strong contenders in the race to build quantum information processing devices in germanium: Ge/Si core/shell nanowires (NWs), Ge hut wires (HWs), and Ge/SiGe planar heterostructures. To advance Ge quantum technology, each of these platforms offers specific advantages to build upon but also poses challenges to overcome, as we shall discuss. Recent critical milestones include: record hole mobility\cite{dobbie_ultra-high_2012,sammak_shallow_2019} up to $10^6$  cm$^2$/Vs; the growth of site-controlled HWs\cite{Gao2020}; the fabrication and measurement of quantum dots in nanowires, hut wires, and quantum wells\cite{huNNano07, hu_hole_2012, Ares2013,Watzinger2016,Watzinger2018,Xu2019,li_coupling_2018,hendrickx_gate-controlled_2018}; the demonstration of electrically driven spin qubits\cite{Watzinger2018}, fast two-qubit logic\cite{hendrickx_fast_2020}, and coherent control, manipulation, and read-out of single hole spins\cite{hendrickx_single-hole_2019}; the demonstration of gate-tunable superconductivity, superconductor-semiconductor hybrids, and coupling of a hole charge to a superconducting resonator\cite{xiangnnano2006,ridderbosAdvMat2018,hendrickx_gate-controlled_2018,vigneau_germanium_2019,li_coupling_2018}.

This Review is organized as follows. We start by introducing the physics of holes in Ge, providing theoretical insights relevant for the use of Ge as a material for qubits based on spin and topological states. We discuss the materials science development in growth and electrical characterisation of Ge/SiGe planar heterostructures, Ge/Si core/shell NWs, and Ge HWs---the three platforms that have enabled building blocks for Ge quantum technology so far. We then summarize the state of experimental progress in Ge devices for quantum technology, providing initial confirmation of theoretical promises. We start with results for quantum dots, including spin states, $g$-factors, charge noise, Pauli spin blockade. We then discuss hole spin qubits experiments, including single-shot spin readout, electrically driven spin qubits, and fast two-qubit logic. Finally, we cover proximity-induced superconductivity in superconductor-semiconductor hybrids, specifically Josephson field-effect transistors, SQUIDs, superconducting quantum point contacts and quantum dots. Based on theoretical considerations and experimental results, we will present our vision for Ge quantum technology towards, for example, macroscopic entanglement between spins, quantum error correction architectures and coherent transfer of quantum information between spin and topological qubits.

\section*{Physics of holes in germanium} 
\label{sec:Physics of holes in germanium}

Only a few years after spins in quantum dots were proposed for quantum computation \cite{Loss1998}, magnetically and electrically driven single-spin rotations and sub-nanosecond square-root-of-SWAP gates
between two spins were demonstrated with conduction-band electrons in lateral GaAs quantum dots \cite{Petta2005, Koppens2006, Nowack2007, KloeffelARCMP2013}. Around the same time, theoretical studies \cite{Bulaev2005, Bulaev2007, Fischer2008, Trif2009} showed that not only filled conduction-band states but also holes, i.e., unfilled valence-band states, are promising candidates for the implementation of spin qubits. The potential of holes for quantum computation was underlined by the encouraging hole-spin lifetimes measured in self-assembled InGaAs quantum dots \cite{Heiss2007, Gerardot2008, Brunner2009, Warburton2013}. By now, holes have been studied in a variety of materials and nanostructures. Below, we focus on holes in Ge-based devices and describe some key insights from a theoretical point of view.

\subsection*{Heavy-hole and light-hole states}
\subsubsection*{Bulk crystal}

In the electronic band structure of bulk Ge (unstrained, diamond cubic lattice), four degenerate valence-band states of highest energy are located at the $\Gamma$ point (Fig.~\ref{theory}a), where the crystal momentum $\hslash \bm{k}$ of the electrons is zero. The fourfold degeneracy results from a spin-orbit interaction at the level of the atoms. More precisely, a term of type $\delta_\textrm{SO} \bm{l} \cdot \bm{s}$ can be derived from the Dirac equation \cite{nolting:book}, where $\hslash \bm{l}$ is the angular momentum associated with the p-type atomic orbitals and $\hslash \bm{s}$ is the bare electron spin. Consequently, an effective spin $\hslash \bm{j} = \hslash (\bm{l} + \bm{s})$ can be defined. From the quantum-mechanical addition of angular momenta and the properties $\bm{l} \cdot \bm{l} = l (l + 1)$ and $\bm{s} \cdot \bm{s} = s (s + 1)$ with $l = 1$ and $s = 1/2$, it follows that $\delta_\textrm{SO} \bm{l} \cdot \bm{s}$ has four degenerate eigenstates with $j = 3/2$ and eigenenergy $\delta_\textrm{SO}/2$ and two degenerate eigenstates with $j = 1/2$ and eigenenergy $- \delta_\textrm{SO}$. The introduced quantum number $j$ is related to the size of the effective spin via $\bm{j} \cdot \bm{j} = j (j + 1)$. The energy difference $\Delta_0 = 3 \delta_\textrm{SO} /2$ is commonly referred to as the spin-orbit gap and separates the topmost valence band ($j = 3/2$) from the spin-orbit split-off band ($j = 1/2$) at the $\Gamma$ point \cite{winkler:book}. For Ge, $\Delta_0 \approx 0.3 \mbox{ eV}$ is relatively large, and so the valence-band states with effective spins 3/2 and 1/2 are energetically more separated than, e.g., for Si, where $\Delta_0$ is only about 44~meV \cite{winkler:book}. 

Near the $\Gamma$ point, the states in the topmost valence band of Ge are well described by the Luttinger-Kohn Hamiltonian\cite{Luttinger1955, Luttinger1956} in the spherical approximation \cite{Winkler2008} 
\begin{equation}
H_\textrm{LK} = - \frac{\hslash^2}{2 m_0} \left[ \left( \gamma_1 + \frac{5}{2} \gamma_s \right) k^2 - 2 \gamma_s \left( \bm{k} \cdot \bm{J} \right)^2 \right] ,
\label{eq:LKHamSphAppr}
\end{equation}
where $k^2 = \bm{k} \cdot \bm{k} = k_x^2 + k_y^2 + k_z^2$, $m_0$ is the free-electron rest mass, $\hslash \bm{J}$ is the operator for an effective spin 3/2, and $\gamma_1$ and $\gamma_s$ are material-dependent parameters. The term $\bm{k} \cdot \bm{J}$ illustrates a crucial feature of hole states: the crystal momentum and the effective spin are closely linked to each other. In particular, the eigenstates of $H_\textrm{LK}$ can be grouped into heavy-hole (HH) and light-hole (LH) states. In the case of HH states, the effective spin is parallel or antiparallel (i.e., spin projections of $\pm 3\hslash/2$) to the hole's direction of motion (given by $\bm{k}$) and the associated eigenenergy is $- \hslash^2 k^2 / 2 m_\textrm{HH}$, where $m_\textrm{HH} = m_0 / (\gamma_1 - 2 \gamma_s)$ is the effective HH mass. In the case of LH states, the spin projection along the direction of motion is only $\pm \hslash/2$ and the eigenenergy is $- \hslash^2 k^2 / 2 m_\textrm{LH}$, where $m_\textrm{LH} = m_0 / (\gamma_1 + 2 \gamma_s)$ is the effective LH mass. For Ge one finds\cite{Lawaetz} $\gamma_1 \approx 13$ and $\gamma_s \approx 5$. Consequently, $m_\textrm{HH}$ is about an order of magnitude greater than $m_\textrm{LH}$. 

\subsubsection*{Planar heterostructures}
The characteristic couplings between the effective spins and the crystal momenta of holes have important consequences when confinement by an external potential is present. In Ge/SiGe planar heterostructures (Fig.~\ref{theory}b), holes are confined to the strained Ge layer because of the valence-band offsets at the interfaces. Since the hole confinement in the out-of-plane direction $z$ is much stronger than in the $x$-$y$ plane, these heterostructures can be considered as quasi-two-dimensional (quasi-2D) systems. Given that the variance of a hole's $z$-coordinate is small, Heisenberg's uncertainty principle implies that the variance of the out-of-plane momentum has a large lower bound. As a consequence, basis states that originate from the topmost valence band are usually classified according to the out-of-plane component $\hslash J_z$ of the effective spin. More precisely, basis states satisfying $J_z = \pm 3/2$ ($J_z = \pm 1/2$) in a quasi-2D structure are commonly referred to as HH (LH) states \cite{winkler:book}. These are related but not identical to the HH and LH states in a bulk semiconductor. 

Recent calculations \cite{Terrazos2018} for realistic Ge/SiGe quantum wells showed that size quantization and strain \cite{birpikus:book} induce a large splitting above 100~meV between states of HH and LH type, with HHs energetically favored. Hence, removing valence-band electrons from quasi-2D Ge systems usually results in hole states whose spin parts consist predominantly of states for which $J_z = \pm 3/2$. Assuming pure HHs \cite{Kesteren1990, winkler:book}, the $k_{x,y}$-dependent terms of Eq.~(\ref{eq:LKHamSphAppr}) simplify to $- \hslash^2 (k_x^2 + k_y^2) / 2 m^\textrm{HH}_\parallel$, where $m^\textrm{HH}_\parallel = m_0 / (\gamma_1 + \gamma_s) \approx 0.055 m_0$ is remarkably light. If $J_z = \pm 1/2$ (pure LHs) instead of $J_z = \pm 3/2$, $m^\textrm{HH}_\parallel$ is replaced by the heavier mass $m^\textrm{LH}_\parallel = m_0 / (\gamma_1 - \gamma_s)$, meaning that HH and LH subbands have reversed characters for the in-plane motion \cite{winkler:book}. Indeed, calculations \cite{Terrazos2018} and experiments with strained, undoped Ge/SiGe quantum wells \cite{lodari_light_2019} resulted in very light in-plane effective masses at the top of the valence band (low-energy holes) of about $0.05 m_0$, whose exact values depend on details such as the strain. Small effective masses lead to large orbital level spacings in quantum dots and are therefore useful, e.g., for the implementation of spin qubits.

\subsubsection*{Nanowires}

\begin{figure}[ht]
\centering
\includegraphics[width=130mm]{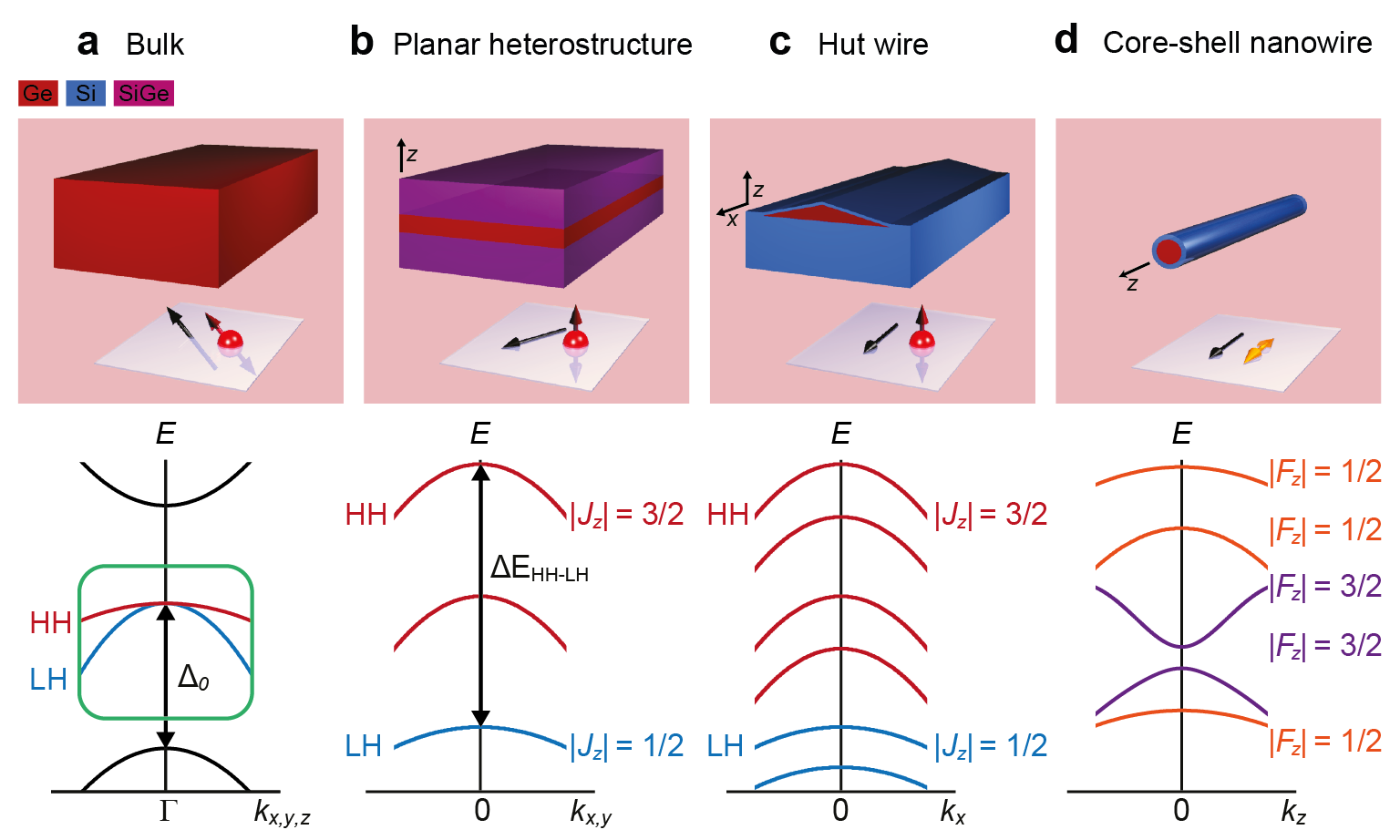}
\caption{\textbf{Quantum confined holes in germanium.} a | In bulk Ge, electrons and holes can move in all directions. An excerpt from the electronic band structure is sketched at the bottom. The HH and LH bands are described by Eq.~(\ref{eq:LKHamSphAppr}), with the spin and momentum satisfying $\bm{J}\cdot\bm{k}/k = \pm 3/2$ for HHs (inset). Corrections beyond the spherical approximation are sometimes relevant \cite{winkler:book, Winkler2008, Kloeffel2018, Terrazos2018} and not considered here for simplicity. In all remaining panels, we zoom in on spectra originating from the bulk HH and LH bands (green square). b | Holes in a Ge/SiGe quantum well can propagate along any in-plane direction. Subbands of HH and LH type arise because of size quantization and strain. Sketched are the two highest HH subbands (red) and the topmost LH subband (blue). The HH-LH splitting is given by $\Delta E_\textrm{HH-LH}$. If corrections are taken into account, the in-plane spectrum becomes anisotropic. c | For Ge HWs, the additional hole confinement along one of the in-plane axes results in additional subbands compared to panel b, a few of which are drawn. Here and in panel b, small HH-LH mixings lead to deviations from $J_z = \pm 3/2$ (insets) and $J_z = \pm 1/2$, respectively, for the subbands of HH and LH type. d | In Ge/Si NWs, the spin states related to $\bm{J}$ are closely linked to the orbital parts of the envelope wave functions and usually strongly mixed. However, due to an approximately cylindrical symmetry, the subbands can be classified according to the total angular momentum $\hslash F_z$ along the wire \cite{Sercel1990, Csontos2009}. The difference $\hslash (F_z - J_z)$ results from orbital angular momenta given by the envelope functions. If cylindrical symmetry is assumed, the topmost subbands satisfy $F_z = \pm 1/2$ (yellow arrow, inset). The sketched spectrum in panel d was adapted with permission from Refs.~\cite{Kloeffel2011, Kloeffel2018}, American Physical Society.}
\label{theory}
\end{figure}

Germanium HWs (Fig.~\ref{theory}c) grown on Si have a triangular cross-section and are fully compressively strained along the in-plane directions $x$ and $y$, where they adopt the Si lattice constant. Given the large width/height aspect ratio of about 10, holes are very strongly confined in the out-of-plane direction $z$. Calculations \cite{Watzinger2016, Gao2020} suggest that low-energy holes in Ge HWs are nearly-pure HHs with a LH admixture below 1\%. Consequently, their effective mass for the motion along the HW is expected to approach the small in-plane mass of low-energy holes in planar heterostructures. 

In Ge/Si core/shell NWs (Fig.~\ref{theory}d), the large valence-band offset of about 500~meV at the Ge-Si interface confines holes to the Ge core. In contrast to the aforementioned systems, there is more than one axis of strongest hole confinement, typically leading to eigenstates that exhibit large HH-LH mixings even at low energies \cite{Sercel1990, Harada2006, Csontos2009}. The motion of low-energy holes along the NW axis $z$ is well described by the term $\hslash^2 k_z^2 / 2 m_\textrm{eff}$, where the effective mass $m_\textrm{eff}$ depends strongly on the strain \cite{Kloeffel2011, Kloeffel2018}. Therefore, $m_\textrm{eff}$ is determined by details of the device, such as the thickness of the Si shell \cite{Menendez2011, Kloeffel2014}. Calculations showed that $m_\textrm{eff}$ can range from small values ($<0.1 m_0$), appealing for spin qubits \cite{Kloeffel2013, Nigg2017}, to infinity and even change sign \cite{Kloeffel2011, Kloeffel2018}. Large effective masses reduce the kinetic energy term and thereby lead to a relative increase of the effects of electron-electron interactions \cite{Maier2014}. This has particularly interesting applications for Majorana physics with holes \cite{Mao2012, Maier2014, Maier2014b}. 

\subsection*{Spin-orbit interaction and g-factors}

The diamond cubic structure has a center of inversion, i.e., Ge is bulk-inversion-symmetric. In contrast, the material interfaces and the external potential in Ge-based devices can induce inversion asymmetry and, consequently, a spin-orbit interaction at the level of the envelope wave functions\cite{winkler:book}. Effects of interface-induced inversion asymmetry on the spin-orbit interaction of electrons and holes have already been studied in detail for a variety of systems \cite{Ivchenko1996, Vervoort1997, Guettler1998, Vervoort1999, Olesberg2001, Hall2003, Golub2004, Nestoklon2008, Prada2011, Furthmeier2016, Wojcik2019}. 

About ten years ago, it was predicted that holes in Ge/Si NWs feature a strong direct Rashba spin-orbit interaction (DRSOI) \cite{Kloeffel2011, Kloeffel2018} if an electric field is applied perpendicular to the wire. The DRSOI requires HH-LH mixing and can be switched on and off via gate electrodes in the experimental setup. In contrast to the well-known case of conduction-band electrons in Rashba NWs, the effective DRSOI coefficient does not rely on couplings between the conduction-band and valence-band states and can therefore be exceptionally large, in agreement with experiments \cite{haoNanoLett2010, Higginbotham2014, Brauns2016a, Wang2017, Sun2018, Vries2018}. For example, spin-orbit energies of a few meV were recently measured for holes in Ge/Si NWs \cite{Wang2017, Sun2018}. According to theoretical studies, a moderate electric field in the Ge core is sufficient for such high spin-orbit energies \cite{Kloeffel2011, Kloeffel2018} and, e.g., allows for efficient hole-spin qubit rotations \cite{Kloeffel2013} via electric dipole spin resonance (EDSR) \cite{Golovach2006}.   

For systems such as Ge/SiGe heterostructures and Ge HWs, in which the hole confinement in one direction clearly exceeds that in the other directions, the DRSOI is less pronounced on account of a large HH-LH splitting \cite{Kloeffel2018, Gao2020}. However, it is known from studying conduction-band electrons in lateral quantum dots \cite{Stano2018, Camenzind2018, Stano2019} that many spin-orbit interaction terms exist which can be of high relevance. For holes in quasi-2D systems, great efforts in both theory and experiment have led to detailed insights and knowledge about the spin-orbit interaction \cite{Winkler2000, winkler:book, Bulaev2007, Winkler2008, Chesi2011, Nichele2014PRL, Nichele2014PRB, Miserev2017, Srinivasan2017, Hung2017, Marcellina2017, Liu2018, Terrazos2018, Gao2020,mizokuchi_hole_2017,moriya_cubic_2014,chou_weak_2018}. Calculations for quantum dots in planar Ge/SiGe heterostructures \cite{Terrazos2018} predicted that the spin-orbit interaction of hole-spin qubits therein can be harnessed to perform rapid (few nanoseconds) qubit rotations by means of EDSR, as measured in experiments\cite{Watzinger2018,hendrickx_fast_2020}.

Many envisaged applications of Ge-based devices require the presence of a magnetic field $\bm{B}$, which we now add to the discussion. Remarkably, the effective $g$ factors of holes (related to the effective Zeeman splittings $|g \mu_B \bm{B}|$) depend strongly on the confinement potential and the magnetic field orientation \cite{Kesteren1990, Nenashev2003, winkler:book, Pryor2006, Winkler2008, Csontos2009, Kloeffel2011, Bree2016, Miserev2017}. For Ge HWs, Ge/SiGe structures, and similar quasi-2D systems, a pure-HH approximation \cite{Kesteren1990} suggests that the effective $g$ factors $g_\perp$ and $g_\parallel$ for an out-of-plane and in-plane $\bm{B}$, respectively, satisfy $|g_\perp| \gg |g_\parallel|$, as usually observed in experiments. \cite{Kesteren1990, katsaros_hybrid_2010, Nichele2014PRL, Watzinger2016, Gao2020}  However, a pure-HH approximation is often insufficient. For example, it turns out that even tiny LH admixtures can substantially reduce the out-of-plane $g$ factor \cite{Wimbauer1994, Ares2013, Durnev2013, Drichko2014, Simion2014, Watzinger2016, Miserev2017}. Nevertheless, we note that a ratio $|g_\perp / g_\parallel|$ of up to 18 was measured for holes in Ge HWs \cite{Watzinger2016}. A particularly attractive property of holes, which is found in various structures, is a strong dependence of the effective $g$ factor on the applied electric field \cite{Pingenot2011, Maier2013, Ares2013, Ares2013APL, Kloeffel2013, Brauns2016a, Marcellina2018, Vries2018}. This property can be used, e.g., to electrically tune the Zeeman splitting and, hence, the resonance condition of hole-spin qubits\cite{hendrickx_single-hole_2019}. Furthermore, it allows for electrically driven spin rotations via $g$-matrix modulation \cite{Pingenot2011, Ares2013APL, Crippa2018, Venitucci2018}.

\subsection*{Hole-spin relaxation, decoherence, and error sources} 

A key criterion for high-quality qubits is that the decoherence time should be much longer than the gate operation times \cite{DiVincenzo2000}. In contrast to III-V materials, Ge and Si contain only a small percentage of atomic nuclei with nonzero spins. Isotopic purification \cite{Itoh1993, AsenPalmer1997, Becker2010, Tyryshkin2012, Veldhorst2014, Muhonen2014, Sigillito2015} even allows for (almost) nuclear-spin-free devices. The contact hyperfine interaction between the remaining nuclear spins and a hole spin is suppressed because the p-type atomic orbitals vanish at the nuclei \cite{Fischer2008, Gerardot2008, Fischer2010, Maier2012, Warburton2013}. It is also worth noting that valley degeneracies, which are present in the lowest conduction bands of Ge and Si, are absent in their topmost valence bands. Thus, it is highly probable that neither valley degrees of freedom \cite{Zwanenburg2013, Vandersypen2019} nor nuclear spins are ultimately a limiting factor for the hole-spin lifetimes in Ge-based devices. 

As a result of the strong spin-orbit interaction and the electric-field-dependent $g$ factors, charge noise is a major source of hole-spin relaxation and decoherence. Fortunately, it is usually possible to identify sweet spots\cite{Burkard1999, Weiss2012, Chesi2014, Wong2015, Reed2016, Martins2016, AbadilloUriel2019} where qubits are to a large extent insensitive to electrical noise. For example, a sweet spot for an idle hole-spin qubit in a Ge/Si NW quantum dot is reached when the effective electric field therein and, hence, the DRSOI of the hole is zero \cite{Kloeffel2013}. Thus, by switching the electric field off unless needed for quantum logic gates, the qubit can be very well isolated from its environment. Phonons \cite{Khaetskii2001, Golovach2004, Stano2006, KloeffelARCMP2013, Zwanenburg2013, Kornich2018, Camenzind2018} are another considerable noise source. However, piezoelectric hole-phonon coupling is suppressed in Ge-based devices because of the bulk inversion symmetry of Si and Ge. Moreover, calculations for hole-spin qubits \cite{Bulaev2005, Trif2009, Maier2013, Li2020} showed how the phonon-mediated qubit decay can be much prolonged, if necessary.

\section*{Material platforms} \label{Materials}

High-quality materials are crucial for viable quantum experiments and quantum technology. In this Section we describe the developments that have established planar heterostructures, NWs, and HWs as the platforms of choice to investigate the properties of holes in Ge that are relevant for quantum information.

\subsection*{Planar heterostructures}

In Ge/SiGe planar heterostructures, the band edge alignment between compressively-strained Ge and relaxed Si$_{1-x}$Ge$_x$ produces a type I band alignment that confines holes in the growth direction creating a two-dimensional hole gas (2DHG)\cite{people_band_1986,people_indirect_1986}. The compressive strain in the epitaxial Ge quantum well is set by the in-plane lattice parameter of the underlying  relaxed Si$_{1-x}$Ge$_x$ buffer layer. In practice, these Si$_{1-x}$Ge$_x$ buffers require a high Ge content ($0.6\leq x \leq0.9$) to support strained Ge epilayers of a reasonable thickness (10 to 30 nm)\cite{Paul2010}, less than the critical thickness for which onset of plastic relaxation takes place\cite{matthews_defects_1976}. The Ge quantum well can be populated with holes from a remote supply layer (modulation doping) or from the ohmic contacts via a gate-induced accumulation in undoped heterostructures. 

\begin{figure}[ht]
\centering
\includegraphics[width=\linewidth]{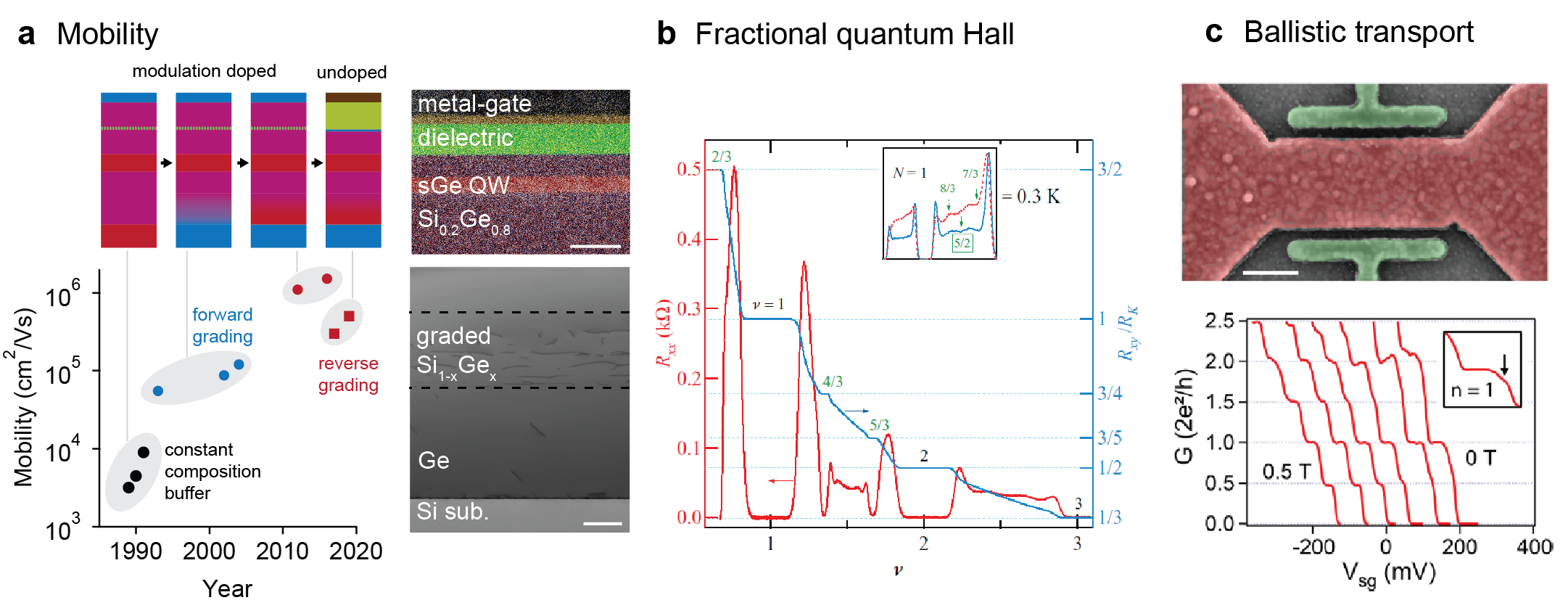}
\caption{\textbf{Planar Ge/SiGe heterostructures} a | Improvement in hole mobility with time (left panel) and evolution towards top-gated undoped Ge quantum wells with high-$\kappa$ dielectric/metal gate stack (top-right panel, scale-bar 50 nm) on reversed-graded relaxed SiGe buffers (bottom-right panel). Images obtained by scanning transmission electron microscopy (bottom-right panel, scale-bar 500 nm) with energy dispersive X-ray analysis (top-right panel). b | Longitudinal resistivity and Hall resistance of a 2D hole gas in Ge/SiGe as a function of filling factors, showing Shubnikov de-Haas minima and quantum Hall plateaus at fractional filling factors. Inset highlights the signature of 5/2 fractional state. c | |Scanning electron microscope image of a gate-defined ballistic 1D channel in planar Ge/SiGe (top panel, scale-bar 200 nm) showing quantization of conductance as a function of gate voltage (bottom panel).  Panel a (right) is adapted with permission from ref~\cite{sammak_shallow_2019}, WILEY-VCH Verlag GmbH \& Co.. Mobility data points are extracted from refs.~\cite{wagner_observation_1989,murakami_high_1990,murakami_strain-controlled_1991} (black circles), refs.~\cite{xie_very_1993,kanel_very_2002,rossner_scattering_2004} (blue circles), refs.~\cite{dobbie_ultra-high_2012,failla_terahertz_2016} (red circles), refs.~\cite{laroche_magneto-transport_2016,sammak_shallow_2019} (red squares). Panel b is adapted with permission from ref~\cite{shi_spinless_2015}, American Physical Society. Panel c is adapted with permission from ref~\cite{mizokuchi_ballistic_2018}, American Chemical Society.}
\label{planar}
\end{figure}

The compressive strain in the Ge quantum well produces large and tunable energy splittings between the HH and LH bands, suggesting that Ge could host a 2D hole gas of very high mobility, since the topmost HH band has a very light effective mass for the in-plane motion (see Sec.~\textit{Physics of holes in germanium}). Motivated by the pursuit of high-mobility $p$-type channels, modulation-doped strained Ge/SiGe heterostructures were pioneered in the 1990s by using molecular beam epitaxy. The highest low-temperature hole mobility (Fig.~\ref{planar}a) improved from less than 10$^4$ cm$^2$/Vs\cite{wagner_observation_1989,murakami_high_1990,murakami_strain-controlled_1991} to 5.5$\times$10$^4$ cm$^2$/Vs\cite{xie_very_1993}, as constant-composition SiGe buffers on Ge substrates were replaced by composition-graded SiGe buffers on Si substrates. 

The progress was modest because several microns of epitaxial growth were required to accommodate the large lattice mismatch between Ge-rich SiGe and the underlying Si substrate, producing increased roughness and high threading dislocation density (TDD $\geq$10$^7$ cm$^{-2}$) during the long-lasting growth. These early efforts are reviewed in ref~\cite{schaffler_high-mobility_1997,lee_strained_2004}. One strategy for improving was to introduce a very steep gradient to Si$_{0.3}$ Ge$_{0.7}$ layers at a fast rate by using low-energy plasma-enhanced chemical vapour deposition\cite{isella_low-energy_2004}. As a result, the mobility in modulation-doped Ge/SiGe was enhanced above 10$^5$  cm$^2$/Vs\cite{kanel_very_2002,rossner_scattering_2004}.   

Later on, the advent of reverse-graded relaxed SiGe buffers\cite{shah_reverse_2008,shah_reverse_2010} with improved TDD ($\sim$10$^6$ cm$^{-2}$) was a turning point in the development of Ge/SiGe heterostructures. High-quality relaxed Ge was grown directly on a Si substrate by reduced-pressure chemical vapor deposition (RP-CVD)\cite{colace_metalsemiconductormetal_1998,gunn2005methods,hartmann_reduced_2004}, followed by reverse grading of the Ge composition in the SiGe buffers. Ultra-high hole mobility exceeding 10$^6$ cm$^2$/Vs was achieved in a modulation-doped strained Ge quantum well at $T <$ 10 K\cite{dobbie_ultra-high_2012}. Key enabler for the one order of magnitude improvement in mobility was the near-perfect epitaxy attainable in an industrial RP-CVD tool, with a very low background of ionized impurities and defects in the Ge quantum well and the surrounding SiGe epilayers.

The exceptional mobility put a spotlight on Ge/SiGe as a platform for quantum devices. Previous studies on Si/SiGe heterostructures showed that undoped structures are better suited for the operation of quantum dots and spin qubits because ionized impurities in the doping layer may cause leakage paths, parasitic channels, and charge noise\cite{lu_enhancement-mode_2011,borselli_pauli_2011,maune_coherent_2012}. This critical know-how was directly ported to the Ge/SiGe platform and efforts were concentrated on undoped accumulation-only heterostructures. As a result the development of quantum devices in planar Ge was greatly accelerated. A high mobility of 5$\times10^5$ cm$^2$/Vs at $T$ = 1.7 K was measured in Ge/SiGe heterostructures field effect transistors (H-FETs)\cite{sammak_shallow_2019} using an industry-compatible high-$\kappa$ dielectric, setting new benchmarks for holes in shallow buried-channel transistors.
Further improvements are expected by an engineered optimization of the stack and device fabrication parameters, such as strain in the quantum well strain, barrier thickness\cite{su_effects_2017,lodari_light_2019}, and dielectric deposition conditions.

The superior quality achieved in reverse-graded Ge/SiGe heterostructures enabled a plethora of quantum transport studies in modulation-doped etched Hall-bar devices\cite{rosner_effective_2003,irisawa_hole_2003,sawano_magnetotransport_2006,sawano_strain_2009,foronda_weak_2014,hassan_anisotropy_2014,moriya_cubic_2014,morrison_observation_2014, failla_narrow_2015,shi_spinless_2015, morrison_complex_2016,holmes_spin-splitting_2016, mironov_fractional_2016,morrison_electronic_2017,drichko_effective_2018,berkutov_quantum_2019} and, more recently, in undoped H-FETs\cite{laroche_magneto-transport_2016,su_effects_2017,lu_density-controlled_2017,lu_effective_2017,chou_weak_2018,hardy_single_2019,lodari_light_2019}. Initial expectations from bandstructure considerations were confirmed and the knowledge-base of confined holes in planar Ge was advanced. Very light in-plane effective masses were measured both in modulation-doped Hall-bars (0.055$m_0$)\cite{morrison_electronic_2017} and in undoped H-FETs  (0.061$m_0$)\cite{lodari_light_2019}, for devices aligned with the ${<}110{>}$ crystallographic direction. The mass is further reduced to 0.035$m_0$ for device alignment along the ${<}100{>}$ direction\cite{morrison_electronic_2017}.
Magnetotransport studies\cite{lu_effective_2017,drichko_effective_2018,sammak_shallow_2019} confirmed that 2D holes in strained-Ge exhibit large out-of-plane effective $g$-factors, with a reported value\cite{lu_effective_2017} $g_{\perp}^* \approx$20 at a density below 1$\times$10$^{11}$ cm$^{-2}$. Large spin-splitting energies (up to $\approx$ 1 meV) were observed\cite{foronda_weak_2014,morrison_observation_2014,moriya_cubic_2014, failla_narrow_2015,holmes_spin-splitting_2016,failla_terahertz_2016,mizokuchi_hole_2017,chou_weak_2018} due to a cubic Rashba-type spin-orbit interaction. Being a single-band system with large Zeeman energy, the 2DHG in high-mobility strained-Ge is an interesting system to investigate fractional quantum Hall physics\cite{shi_spinless_2015, mironov_fractional_2016,berkutov_quantum_2019} (Fig.~\ref{planar}b), with fractional states observed even above 4 K. Furthermore, the ability to control the carrier density within a single H-FET has given insights into the percolation threshold density to establish a metallic 2D channel in Ge. A very low value of $\sim2\times10^{10}$ cm$^{-2}$ was reported\cite{lodari2020}, indicative of very low disorder in the low density regime relevant for spin qubits in quantum dots.

One dimensional hole channels and quantum point contacts were the first proof-of-principle devices fabricated by further electrostatic confinement of 2DHGs in undoped strained Ge/SiGe heterostructures \cite{gul_quantum_2017,mizokuchi_ballistic_2018,gul_self-organised_2018}. The high quality of the originating heterostructure, with hole mean free paths in the order of a micron, enabled the observation of quantum ballistic transport (Fig.~\ref{planar}c) in one-dimensional channels as long as 600 nm\cite{mizokuchi_ballistic_2018}, clean quantized conductance plateaus\cite{gul_quantum_2017,mizokuchi_ballistic_2018}, and possibly charge fractionalisation due to strong interactions at low densities\cite{gul_self-organised_2018}. Furthermore, a large $g$-factor anisotropy was observed ($g_{\perp}>$10 and $g_{\parallel}<$1), confirming the heavy-hole character of the dominant valence band states, imparted by the strong vertical confinement in the heterostructure growth direction.

\subsection*{Nanowires}
\subsubsection*{Ge/Si core/shell nanowires}
The Ge/Si core/shell NW heterostructure has a type II band alignment with a valence band offset of about 500 meV where holes are strongly confined in the Ge core\cite{lauhonNature02}. The development of Ge/Si core/shell NWs  traces back to the growth of Si and Ge NWs by the vapor-liquid-solid (VLS) mechanism using  laser-ablated nanometer-diameter clusters of Au as catalyst\cite{moralesScience98}. The NWs can grow up to 30 $\mu$m in length, with the metal catalysts defining precisely the position of the NWs and their diameter, in the range of 6-100 nm and 3-100 nm for Si and Ge NWs, respectively.  By depositing homogeneously sized gold catalysts onto oxide-free Si, vertically <111> oriented Ge NWs of uniform diameter and length on Si(111) substrates were obtained\cite{Woodruff2007}. Ordered arrays of Ge NWs with controlled position were demonstrated by using lithographically patterned Au nanoparticles\cite{dayeh_direct_2010} and kinked Ge NW structures were reported with iterative control over the nucleation and growth of NWs\cite{tian_single-crystalline_2009}. The NW growth rate decreases for smaller diameters, with a cutoff at sufficiently small sizes\cite{dayeh_direct_2010}. Furthermore, the NW growth orientation was strongly correlated with the NW diameter. Ge NWs with diameter larger than 20 nm are mostly oriented along the <111> direction, while <110> and <112>  orientations are preferred for smaller diameter NWs\cite{goldthorpe_inhibiting_2009,Conesa-BojNanolett2017}. 

\begin{figure}[ht]
\centering
\includegraphics[width=\linewidth]{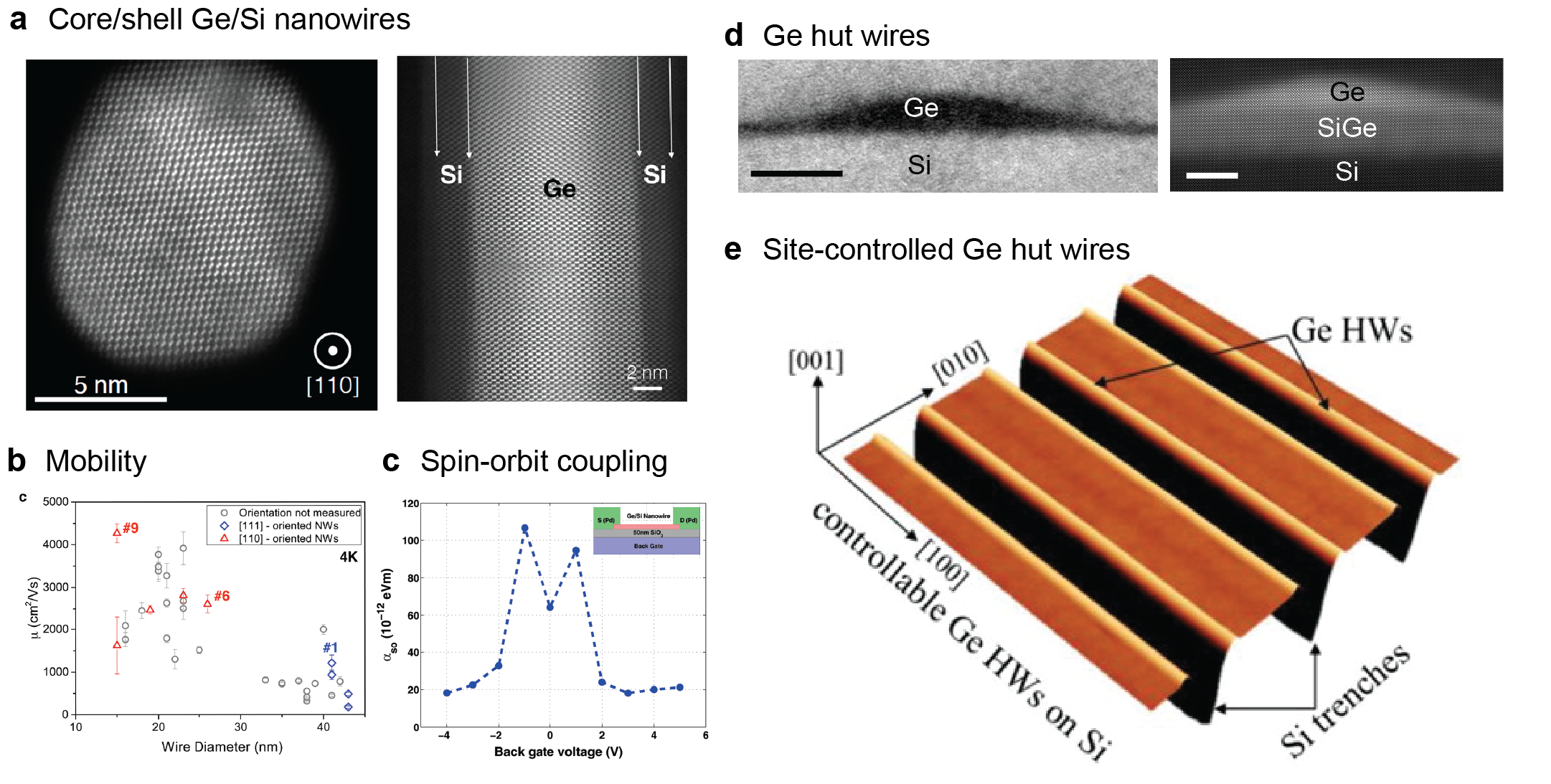}
\caption{\textbf{Ge-based nanowires} a | High-angle annular dark-field image of a Ge/Si core/shell NW along the radial (left panel) and axial (right panel) direction. b | Hole mobility of NWs as a function of wire diameter for different NW orientation. c | Spin-orbit coupling strength $\alpha_{so}$ as a function of the back gate voltage in a NW field effect transistor (inset) d |  Transmission electron microscope image of a Ge hut wire on Si covered with Si cap (left panel) and image of a Ge HW sitting on SiGe (right panel) e | Atomic force microscopy image of site-controlled Ge hut wires with highly uniform size on trench-patterned Si. Panel a and b are adapted with permission from ref \cite{Conesa-BojNanolett2017}, American Chemical Society. Panel c is adapted with permission from ref \cite{haoNanoLett2010}, American Chemical Society. Panel d is adapted with permission from ref~\cite{Zhang2012}, American Physical Society and from ref~\cite{Gao2020}, WILEY-VCH Verlag GmbH \& Co.. Panel e is adapted with permission from ref~\cite{Gao2020}, WILEY-VCH Verlag GmbH \& Co. }
\label{nanowires}
\end{figure}

The know-how accumulated in the development of Si and Ge NWs paved the way towards the epitaxial growth of core/shell Ge/Si NW heterostructures , which were synthesized by the vapor-phase deposition of a conformal Si shell on the VLS-grown Ge core\cite{lauhonNature02}. The mostly amorphous Si shell was grown at low temperature to form an abrupt Ge/Si interface and was then fully crystallized by \textit{in-situ} thermal annealing with negligible intermixing. With elevated growth temperature and the introduction of HCl to avoid strain-driven surface roughing during the shell growth, high quality Ge/Si NWs (Fig.~\ref{nanowires}a) with low density of defects and sharp interface were realized \cite{goldthorpe_inhibiting_2009,Conesa-BojNanolett2017}. In addition, radial multi-shell doping and even modulation doping core/shell NWs have been demonstrated\cite{lauhonNature02,dillen_radial_2014}. Recently, an advanced radial heterostructure was synthesized, with a core made of short segments of crystalline Ge and aluminum surrounded by a Si shell (Ge-Al/Si core/shell NWs), providing avenues for the synthesis of epitaxial superconductor-semiconductor heterostructures in Ge NWs\cite{sistani_highly_2019}.

Shortly after the demonstration of Ge/Si core/shell NWs, in 2005 a high-quality one-dimensional hole gas (1DHG) in Ge was demonstrated\cite{LuPNAS05}. Controlled single hole transistor behaviour was observed at low temperature when the Si shell was used as a tunnel barrier to the Ge core. Enhanced mobility in undoped structures resulted in carrier mean-free-path with possible signatures of ballistic transport up to a few hundreds nanometers, both at low and room temperatures\cite{LuPNAS05,kotekar-patil_quasiballistic_2017}. Further studies revealed also a dependence of the carrier mobility with the NW orientation (Fig.~\ref{nanowires}b). Compared to the conventional <111>-oriented NWs, <110>-oriented Ge/Si core/shell NWs show substantially enhanced hole mobility with values as high as 4200 cm$^2$/Vs at 4 K\cite{Conesa-BojNanolett2017}.

The high-quality achieved in Ge/Si core/shell NWs allowed for a detailed investigation of quantum  transport of 1DHGs at low temperatures. In ref~\cite{haoNanoLett2010} weak antilocalization was observed due to spin-orbit coupling that could be modulated by a factor of up to $\approx \times$5 with an external electric field (Fig.~\ref{nanowires}c). More recently, detailed measurements of spin-orbit energies in 1DHGs Ge/Si NWs\cite{Wang2017,Sun2018} confirmed the large values expected by the DRSOI\cite{Kloeffel2011,Kloeffel2018}. In ref~\cite{Wang2017} highly tunable 1DHG FET devices with a dual-gate architecture were fabricated and a very large and tunable spin–orbit energy was evaluated in the range 1-6 meV, almost one order of magnitude larger than that reported for NWs in III-Vs semiconductors. A comparable a spin-orbit energy of 2.1 meV was evaluated in Ref. \cite{Sun2018}, and a $g$-factor of 3.6 was measured due to strong HH-LH mixing in the wires. 

Despite the tremendous progress over the past 15 years, there are still two key material challenges to address for further advancing 1DHGs in Ge/Si NWs. One is the development of catalyst-free growth of high quality Ge NWs, since the use of metal catalyst inevitably introduces metal contamination in the semiconductor NWs. The other challenge is how to transfer and arrange the out-of-plane NWs in large scale into in-plane with fully controllable positions. Preliminary results in this direction demonstrated the controlled catalyst-free dormation and a 1DHG in Ge by fabricating Si/Ge/Si core-double shell nanowires using a combination of nanoimprint lithography, etching, and chemical vapor deposition\cite{zhang_controlling_2019}.

\subsubsection*{Ge hut wires}
The growth of Ge HWs (Fig.~\ref{nanowires}d) was enlightened by the epitaxial growth of self-assembled Ge quantum dots (QDs) on Si, studied extensively in the 1990s\cite{Tersoff}. Three-dimensional (3D) self-assembled Ge QDs\cite{Mo1990} were demonstrated after the deposition on a Si(001) substrate of about 4 monolayer (ML) of strained Ge. Such Ge QDs were called “hut clusters” and had a height of about 1-2 nm and four \{105\} facets with a rectangular base area. \textit{In-situ} scanning tunneling microscopy studies\cite{McKay} showed that hut clusters grow slowly at a decreasing rate throughout the anneal. In 2012 it was reported that the hut clusters elongate along the in-plane [001] or [010] crystallographic directions into micrometers long Ge HWs (Fig.~\ref{nanowires}d) under appropriate molecular beam epitaxy growth and anneal conditions\cite{Zhang2012}. 

The in-plane Ge HWs are uniform with an average height of about 2 nm. They have the hut cluster’s characteristic of four \{105\} facets and a shallow triangular cross-section with side inclination angle of 11.3$^{\circ}$. This cross-section is completely different from the VLS grown cilindric NWs, and resembles more that one of a laterally confined quasi-2D system.  The planar density of HWs can be controlled simply by the amount of the initially deposited Ge. A slightly increased Ge amount leads to a high density of short HWs. In contrast, a decreased Ge amount results in the reduction of nucleation rate\cite{tersoff_competing_1994} and correspondingly a low density and long HWs up to $\approx$ 2 micrometers, with a large length-to-height aspect ratio  up to $\approx$1000. 

The Ge HWs have a constant height and width along the wire on the same atomic terrace. Their height increases or decreases one atomic layer thickness when the wire crosses an atomic step. By choosing the morphology of the underlying Si surface, homogeneous or tapered Ge HW can be obtained\cite{Zhang2012,Watzinger2014}. The further deposition of a few-nm-thick epitaxial layer of Si at low-temperature (350$^{\circ}$C) is sufficient to fully strain the Ge HWs and achieve a sharp Si/Ge interface. Transistor-type devices made from single wires support large current densities (10$^7$A/cm$^2$) at low temperature\cite{Zhang2012}. The hole channel could be pinched off and, near pinch-off, single-hole transport was observed due to Coulomb blockade at temperatures below a few K. Mobility and quantum transport studies in single Ge HWs are very limited compared to 2DHGs in Ge/SiGe and 1DHGs in Ge/Si NWs.

Although growth of uniform in-plane Ge HWs is catalyst-free, the random position of the HWs on the Si substrate is a bottleneck for the scale up of quantum devices. Recent work has addressed this challenge by exploiting the self-assembled Ge growth on a lithographically patterned Si substrate. Site-controlled growth of highly uniform in-plane Ge HWs on Si (001) substrates (Fig.~\ref{nanowires}d) has been successfully demonstrated\cite{Gao2020}. The Ge HWs grow selectively on an initially formed 1D SiGe layer at the <100>-oriented trench edges of patterned Si (Fig.~\ref{nanowires}e). They have a height of about 3.8 nm with a standard deviation of merely 0.11 nm, and their position, distance, length can be precisely controlled. As shown in (Fig.~\ref{nanowires}d), sharp Ge/SiGe and Si/Ge interfaces are observed. In addition, closely-spaced parallel HWs and even square-shaped or L-shaped structures consisting of four or two Ge HWs can be obtained.  

\section*{Germanium quantum devices} \label{Devices}
Core/shell NWs, HWs, and planar heterostructures have emerged in the past fifteen years as the material platforms of choice to demonstrate the basic building blocks for Ge-based quantum information processing. Results on quantum dots, spin qubits, and superconductor-semiconductor hybrids bode well towards a fast and scalable Ge quantum technology that can possibly rely on hybrid combinations of semiconductor, superconductor and topological phases.

\begin{figure}[ht!]
\centering
\includegraphics[width=158mm]{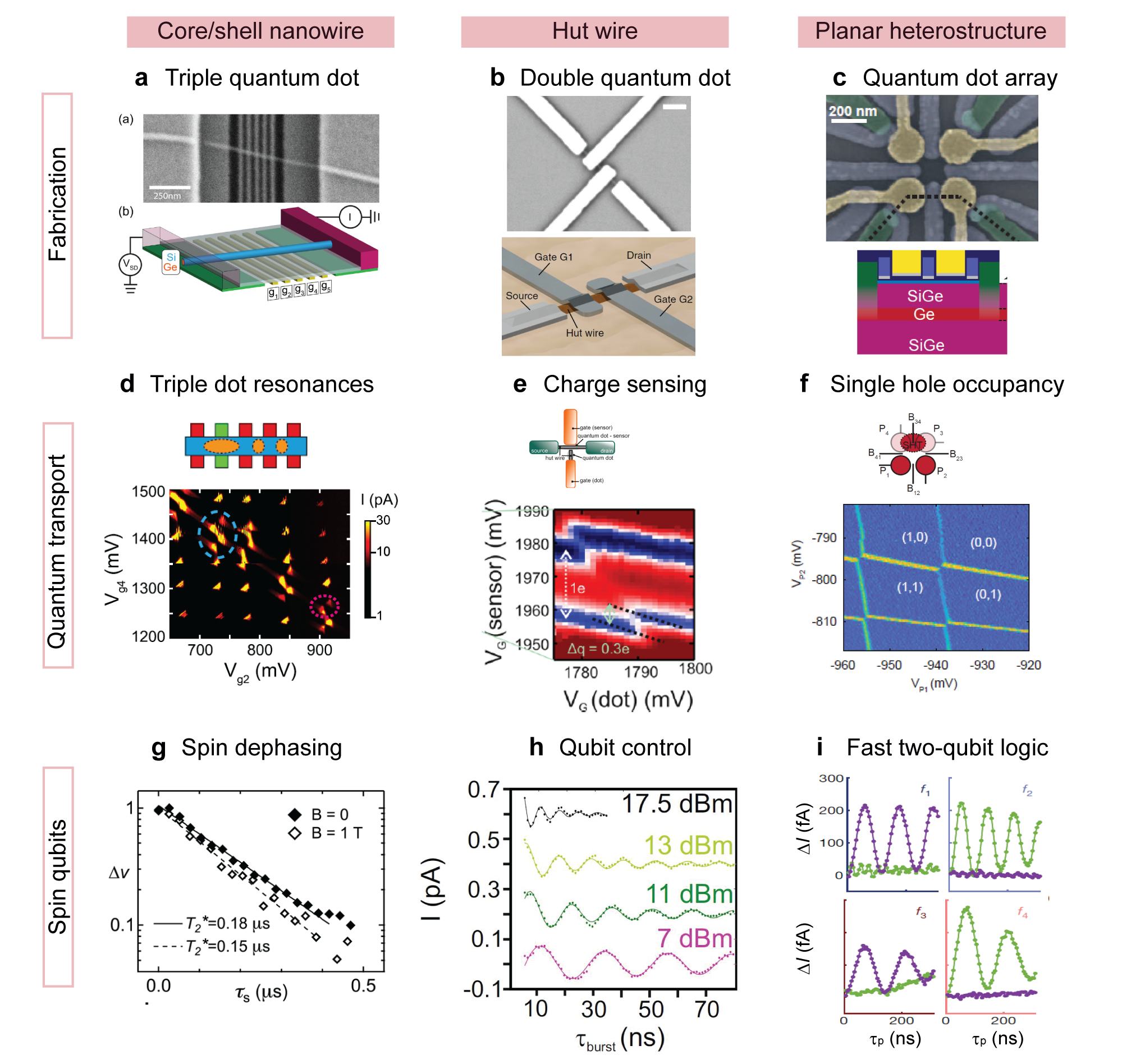}
\caption{\textbf{State of the art quantum dots and qubits in core/shell nanowires, hut wires and planar systems.} a | Triple quantum dots have been realized in nanowires, b | double quantum dots in hut wires, and c | quadruple quantum dots in planar systems. d | Triple dot stability diagram of a nanowire measured in transport. e | Charge sensing of a single quantum dot in a hut wire. f | Charge sensing single-hole occupancy in a reconfigurable quadruple quantum dot in planar Ge. g | Fast pulsing in nanowires enabled to measured spin relaxation and coherence. h | EDSR enabled fast driving in Ge HWs with Rabi frequencies approaching 150 MHz. i | By combining single and two qubit gates, universal quantum logic is demonstrated in planar Ge. EDSR driving enables to execute CROT operation with CX-gates demonstrated within 55ns. Panel a and d are adapted with permission from ref~\cite{froningAPL2018}, American Institute of Physics. Panel b (scale bar is 200 nm) and h are adapted with permission from ref~\cite{Watzinger2018}, Springer Nature Limited, with inclusion of additional data (panel h, black curve 17.5 dBm). Panel c and f are adapted with permission from ref~\cite{lawrie_quantum_2020}, American Institute of Physics. Panel e is adapted with permission from ref~\cite{Vukusic2017}, American Chemical Society. Panel g is adapted with permission from ref~\cite{higginbotham_hole_2014}, American Chemical Society.  Panel i is adapted with permission from ref~\cite{hendrickx_fast_2020}, Springer Nature Limited.}
\label{qubits}
\end{figure}

\subsection*{Quantum dots}
Low-dimensional systems such as planar heterostructures and nanowires are naturally suited for quantum dot experiments. In these systems the geometrical confinement to one or two dimensions requires additional confinement to create quantum dots, i.e. 0-dimensional islands. Gate electrodes isolated via suitable dielectrics and/or Schottky barriers can be used for the tunnel barriers defining the quantum dot connected to source, drain and gate electrodes. The experimental demonstrations of gate-controlled Ge quantum dots have allowed measurements of spin states, $g$-factors and Pauli spin blockade. This knowledge is typically gained in magnetotransport or charge sensing measurements at low temperatures. 

\subsubsection*{Nanowires}
The first report of a 1DHG in Ge/Si core/shell NWs included the observation of Coulomb blockade oscillations due to single hole tunneling from the contacts through the thin Si shell into the Ge core\cite{LuPNAS05}. Regular Coulomb diamonds were observed indicating low disorder in the wire. Such wires were then combined with local gate electrodes to create double quantum dots and integrated charge sensors \cite{huNNano07}. These experiments demonstrated the possibility to control the tunnel coupling, albeit in the multi-hole regime. Further advances in material quality \cite{Conesa-BojNanolett2017} enabled quantum dots with an increased tuning capability\cite{BraunsAPL2016} allowing to define single, double and triple QDs in Ge/Si NWs (Fig.~\ref{qubits}a), with a low hole occupation number\cite{froningAPL2018} (Fig.~\ref{qubits}d). Transport measurements in nanowire quantum dots have also been used to probe the spin states of holes and for strongly confined quantum dots a $g$-factor $\sim 2$ has been measured \cite{roddaroPRL08}. Nonetheless, a strong anisotropy was predicted\cite{Maier2013} and measured $g^*_{\textrm{max}}/g^*_{\textrm{min}} \approx 13$ in nanowire quantum dots \cite{Brauns2016a}. This large anisotropy follows from an almost vanishing $g$-factor ($g^*_{min}$ = 0.2) along the NW axis. Pauli spin blockade can elucidate the spin-flip mechanisms through the behaviour of the leakage current as a function of magnetic field \cite{Brauns2016b, ZarassiPRB2017, froningAPL2018}.

\subsubsection*{Hut wires}
Similarly to the Ge/Si NWs, the first report on Ge HWs also included the observation of single hole transport\cite{Zhang2012}. However, the Coulomb diamonds observed in the HWs were more irregular compared to the Ge/Si due to the increased disorder in this system\cite{Zhang2012}. Subsequent improvements in materials and fabrication processes demonstrated also in this system gate-controlled single and double quantum dots\cite{Watzinger2016,Vukusic2018,Watzinger2018} (Fig.~\ref{qubits}b) and rf charge sensing\cite{Vukusic2017} (Fig.~\ref{qubits}b). Spin states and $g$-factors can be measured in quantum dots via ground-state and excited-state magnetospectroscopy \cite{Hanson2007}. The hole band structure often leads to anisotropic $g$-factors. In hut wires, the lowest energy states are mostly of heavy-hole nature and large anisotropies of $\approx 16-18$ with heavy hole $g$-factors up to 4.4 were found in HWs quantum dots\cite{Watzinger2016}.   

\subsubsection*{Planar quantum dots}
After high-quality undoped Ge/SiGe heterostructures \cite{sammak_shallow_2019} became available, remarkable development has been made with planar quantum dots and in only two years device technology matured from demonstrations of gate-controlled single quantum dots \cite{hendrickx_gate-controlled_2018} to quadruple quantum dots arranged in a two-dimensional array\cite{lawrie_quantum_2020} (Fig.~\ref{qubits}c), highlighting the beneficial aspects of planar systems for scalability. Owing to the high mobility \cite{sammak_shallow_2019} and light mass of holes \cite{lodari_light_2019}, comparatively large quantum dots (diameter of $\approx$ 100 nm) can be defined \cite{hendrickx_gate-controlled_2018} and tuned to contain only a single hole \cite{hendrickx_single-hole_2019, lawrie_quantum_2020} in Ge/SiGe quantum wells. Furthermore, these devices are compatible with electric gate fabrication using single-layer technology \cite{hendrickx_gate-controlled_2018}. Moreover, these quantum dots can be directly contacted by aluminum leads, resulting in a completely dopant-free architecture\cite{hendrickx_gate-controlled_2018}. The low disorder of the originating heterostructures in quantum dots are electrically stable and show excellent homogeneity together with limited charge noise, $1 \mu$eV$/\sqrt{\textrm{Hz}}$ at 1 Hz \cite{hendrickx_gate-controlled_2018} and double quantum dots have been reported \cite{hendrickx_fast_2020,hardy_single_2019,hofmann2019assessing}. In-plane $g$-factors $\sim$ 0.2-0.3 were measured consistently in quantum dots in planar Ge\cite{hendrickx_gate-controlled_2018,hendrickx_fast_2020,hofmann2019assessing}, together with an anisotropy $g_{out}/g_{in} \approx$ 18. This large anisotropy denotes the HH character of the first subbands, as expected in the limit of strong confinement along the growth axis. In planar Ge quantum dots, a remarkable control over the inter-dot tunnel coupling and detuning has been obtained \cite{hendrickx_fast_2020}, enabling to operate at the charge symmetry point for optimized quantum control \cite{hendrickx_fast_2020}. Furthermore, single hole occupation has been achieved \cite{lawrie_quantum_2020, hendrickx_single-hole_2019} and in quadruple quantum dots\cite{lawrie_quantum_2020}(Fig.~\ref{qubits}f), rf charge sensing was realized and fully functional two-dimensional arrays were operated. 

\subsection*{Spin qubits}
\begin{figure}[ht!]
\centering
\includegraphics[width=120mm]{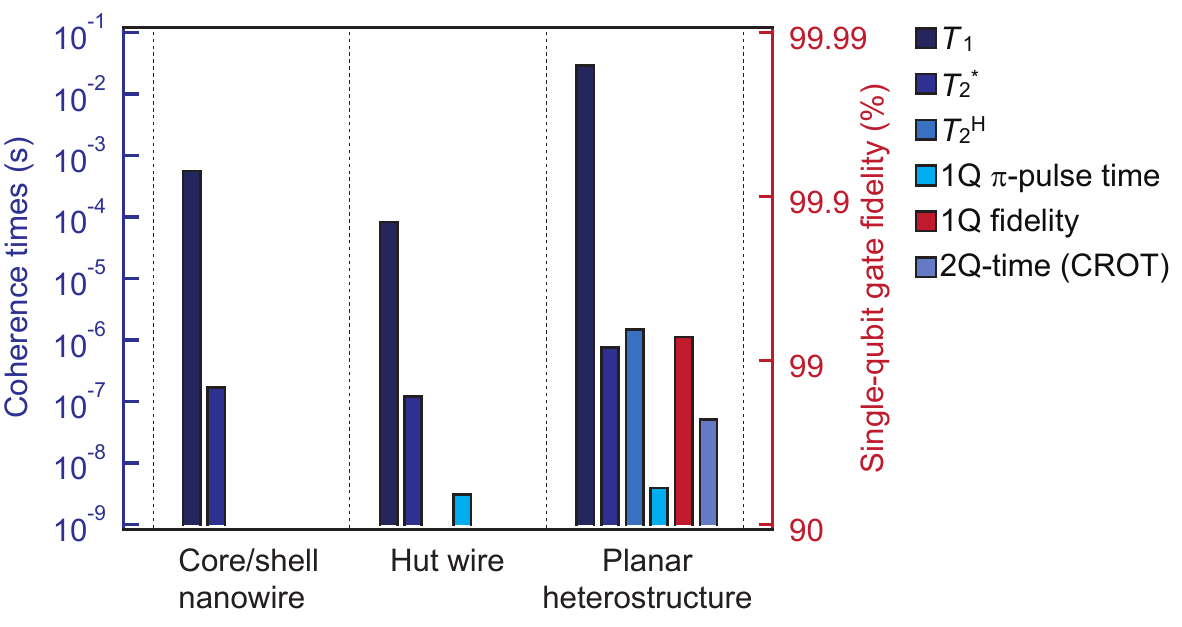}
\caption{\textbf{State of the art results in coherence times and single-qubit gate fidelity for core/shell nanowires, hut wires, and planar heterostructures.} Spin relaxation and dephasing times $T_1$ and $T_2^*$ have been measured in each platform. Coherent Rabi rotation with frequencies above 100 MHz have been performed with hut wires and planar structures. Extended coherence times have been realized in Ge using dynamical decoupling, fidelities nearing fault-tolerant thresholds have been measured using randomized benchmarking. Fast two-qubit logic has been executed using exchange coupled quantum dots in planar Ge. Data for nanowires is adapted with permission from ref~\cite{hu_hole_2012}, Springer Nature Limited, and ref~\cite{higginbotham_hole_2014}, American Chemical Society. Data for hut wires is adapted with permission from ref~\cite{Vukusic2018}, American Chemical Society, and ref~\cite{Watzinger2018}, Springer Nature Limited. Data for planar heterostructures is adapted with permission from ref~\cite{lawrieinprep},and ref~\cite{hendrickx_fast_2020}, Springer Nature Limited.}
\label{timescales}
\end{figure}

Spin relaxation and dephasing has been measured in Ge/Si core/shell NWs\cite{higginbotham_hole_2014}, single qubit rotations have been performed in Ge HWs \cite{Watzinger2018}, while single qubit operation with a single hole spin\cite{hendrickx_single-hole_2019} and two-qubit logic has been executed in planar structures\cite{hendrickx_fast_2020} (see Fig.~\ref{timescales}). The key challenge for quantum information is to increase the number of qubits and to scale in two dimensions. In other systems like electrons in Si, coherent control of single spins require large objects such as microwave striplines or nanomagnets and it is an open question how to integrate this with a larger number of exchange coupled qubits \cite{vandersypen_interfacing_2017}. The spin-orbit coupling for holes in Ge offers an opportunity to circumvent this bottleneck and hence facilitate scalability. Nonetheless, it also opens new sources for decoherence and in this part we review quantum coherence and provide perspective for future experiments.

The spin lifetime and coherence can be obtained by exploiting the exchange interaction in a double quantum dot to prepare suitable spin states (e.g. spin singlets or triplets), or by Rabi driving the spin to the desired states using microwave pulses and monitoring the time evolution. Spin lifetimes have been measured in Ge nanowires using fast gate pulsing and tuning the exchange interaction (Fig.~\ref{qubits}g). A spin lifetime $T_1 = 0.6$ ms has been measured in the zero-magnetic-field limit, which reduces to $T_1 = 0.2$ ms at 1 T \cite{hu_hole_2012}. These results have been obtained using transport measurements for an unknown number of confined holes estimated between 10 and 50. By integrating charge sensors and barrier gates to control the tunnel rates, isolated single spins in Ge can be measured, which can have spin lifetimes $T_1 = 32$ ms \cite{hendrickx_single-hole_2019, lawrieinprep} in a magnetic field $B = 0.67$ T for planar quantum dots. Ramsey sequences have been used to measure the spin dephasing times, yielding $T_2^*$ = $0.18$ $\mu$s for Ge/Si core/shell NWs\cite{higginbotham_hole_2014}, $T_2^*$ = $0.13$ $\mu s$ for Ge HWs\cite{Watzinger2018}, and $T_2^*$ = $0.82$ $\mu s$ for planar Ge \cite{hendrickx_fast_2020, hendrickx_single-hole_2019}. Interestingly, the measured dephasing time in planar Ge significantly exceeds $T_2^*$ = 270 ns reported for holes in Si \cite{hutin_si_2018}. Furthermore, in planar Ge the coherence could be extended by using dynamical decoupling and a Hahn echo $T_2^H = 1.9$ $\mu s$ has been reported \cite{hendrickx_fast_2020}.

Holes in Ge can be all-electrically controlled by electric dipole spin resonance (EDSR) \cite{Bulaev2005, Bulaev2007} and this has been demonstrated in Ge HWs \cite{Watzinger2018}  and Ge planar structures\cite{hendrickx_fast_2020, hendrickx_single-hole_2019}. Fast Rabi driving speeds exceeding 100 MHz have been obtained by simply applying electric tones to the same gates that are used to define the quantum dots \cite{Watzinger2018, hendrickx_fast_2020}. The first Ge qubit was realized in Ge HWs in a double quantum dot and readout was performed using Pauli spin blockade in transport\cite{Watzinger2018}. By fixing the gate voltages to the Pauli spin blockade regime and by sweeping the microwave frequency while stepping the magnetic field an increase in the current could be observed when the condition $hf = g\mu_B B$ is fulfilled (Fig.~\ref{qubits}h). This current increase is due to the fact that the microwave field leads to spin rotation via EDSR. From the slope of the EDSR line the $g$-factor can be extracted and its anisotropy with the magnetic field can be studied. In the case of different $g$-factor values in the two QDs two lines can be observed \cite{hendrickx_fast_2020}. To demonstrate coherent qubit rotations, the quantum dot system is pulsed to Coulomb blockade, where a short rf burst is applied, and pulsed back to the Pauli spin blockade regime for readout. By changing the duration of the rf burst, coherent spin oscillations can be realized (Fig.~\ref{qubits})h). A particular challenge in transport measurements is that the current amplitude depends on the time scale of the experiment, $\delta I=2e/t_{cycle}$. For typical cycle times $t_{cycle} \approx 1$ $\mu$s, the resulting current is only a fraction of a pA. In order to obtain high quality data, lock-in techniques have been used together with reference pulses to compensate for slow variations in the offset current \cite{hendrickx_fast_2020}.

By coherently controlling both spins in a planar Ge double quantum dot, the two-qubit Hilbert space can be assessed by combining single qubit rotations and a two-qubit CROT gate\cite{hendrickx_fast_2020} (Fig.~\ref{qubits}i). The individual qubit fidelities have been measured using randomized benchmarking \cite{knill_randomized_2008}. In this protocol, the fidelity decay is monitored as a function of a series of randomly drawn Cliffords. An average single qubit fidelity $F_C = 99.3 \%$ has been observed. The two-qubit logic is then obtained by tuning the exchange interaction. This can be done via the detuning energy in the presence of finite tunnel coupling or by directly controlling the tunnel coupling. The latter method has proven advantageous as it allows to operate at the charge symmetry point, where the system is to first order insensitive to detuning noise\cite{bertrand_quantum_2015,Reed2016,Martins2016}. This becomes evident from the strong dependence of quantum coherence on detuning \cite{hendrickx_fast_2020}. This suggests significant detuning noise, most likely due to charge noise. Here, the advantage of low disorder and low effective mass in planar Ge becomes evident, as it allows to incorporate a dedicated barrier gate to tune to the charge symmetry point while having significant exchange interaction. In this regime, algorithms have been performed to demonstrate the coherence of the two-qubit logic with CROT operations executed in timescales of only 55 ns\cite{hendrickx_fast_2020}. 

\subsection*{Superconductor-semiconductor hybrids}
In superconductor(S)-semiconductor(Sm) hybrids\cite{ de_franceschi_hybrid_2010, lutchyn_majorana_2018} the semiconducting component of the device acquires superconducting properties from the superconductor, a phenomenon known as the superconducting proximity effect. As a result, a device with a short semiconductor section connecting two superconducting electrodes can behave as a superconducting Josephson junction whose critical current $I_c$ is tuned by a gate voltage. Such a device, known as Josephson field-effect transistor \cite{clark_feasibility_1980} (JoFET), was integrated recently in a superconducting ‘transmon’ qubit enabling an electrostatic control of the qubit energy level spacing \cite{de_lange_realization_2015,larsen_semiconductor-nanowire-based_2015, casparis_superconducting_2018}. Gatemons are of practical interest for large-scale integration, since their tuning does not require mA-scale currents flowing through inductively-coupled circuits. Hybrid S-Sm structures are also highly relevant for topological quantum computation based on MZM. These exotic quasiparticle states have been predicted to emerge at the edges of a one-dimensional S-Sm system provided the semiconductor has strong spin-orbit coupling and sufficiently large magnetic field is applied along certain directions\cite{lutchyn_majorana_2018}. 

Most of the experimental research has been focusing on hybrid S-Sm systems involving small band-gap III-V semiconductors such as InAs and InSb. Germanium is an appealing alternative semiconductor to host hybrid S-Sm devices and their large scale integration on Si. In Ge-based S-Sm hybrids, the favored semiconductor quasiparticle states are holes and not electrons. In fact, multiple experiments have shown that the Fermi energy of certain superconducting metals can pin close to the valence band edge of Ge resulting in a high-transparency S-Sm interface. 

\begin{figure}[ht!]
\centering
\includegraphics[width=160MM]{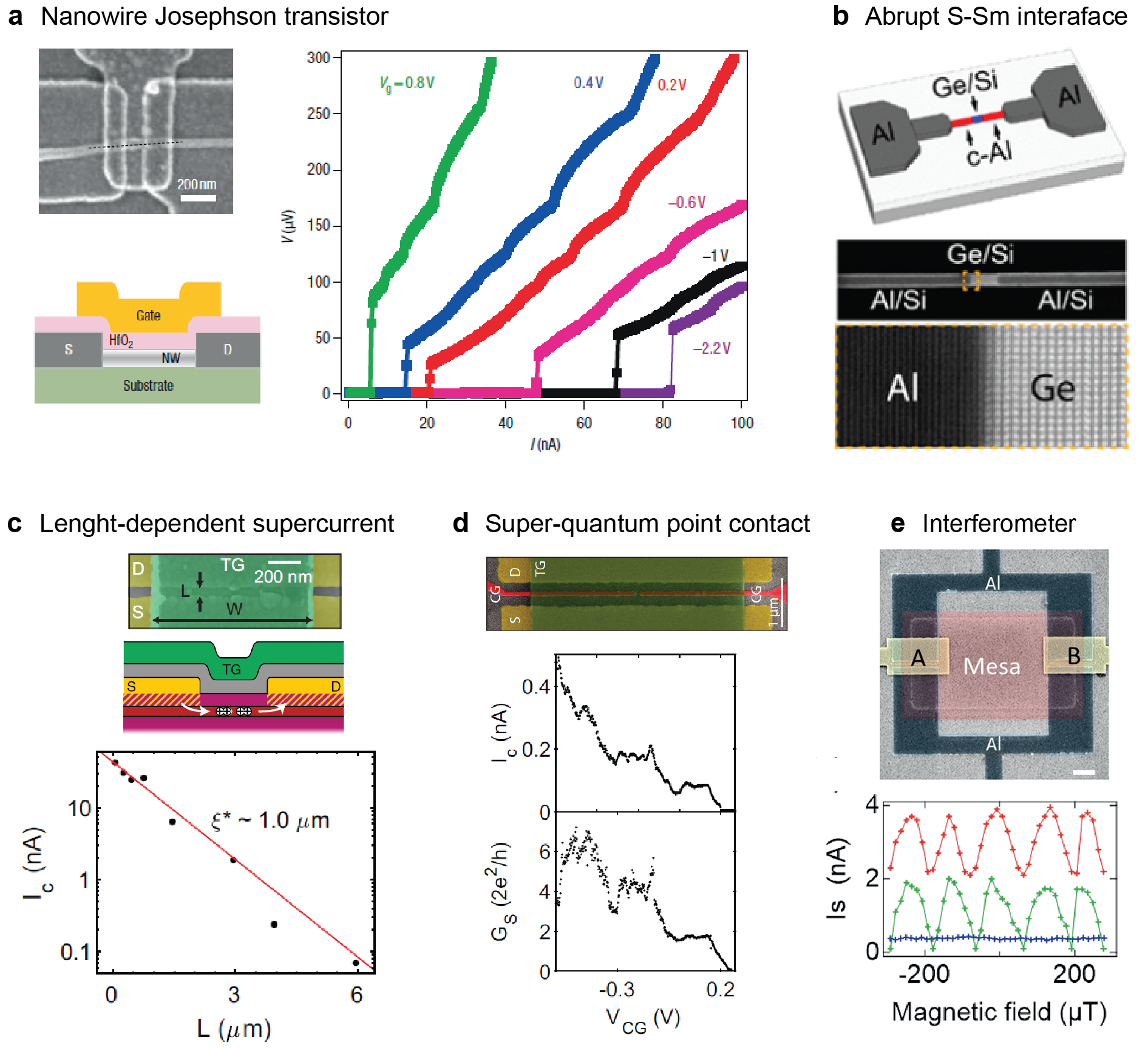}
\caption{\textbf{Superconductor-semiconductor hybrids in Ge nanowires and planar systems} Josephson junctions have been realized in a | Ge/Si core/shell NWs and c | planar Ge/SiGe. In these junctions, the switching current---denoting the transition from superconducting to normal state---is gate dependent (see current-voltage traces at 60 mK). b | Atomically abrupt superconductor-semiconductor interfaces have been engineered in axial Al-Ge-Al NW heterostructure with an ultrathin Si shell wrapped around it. c-e | In planar systems supercurrent transport has been observed over $\mu$m-long channels, and superconducting quantum point contacts and interferometers have been fabricated. c |Length dependence of the supercurrent in planar systems. A purely exponential decay is observed with a decay length of 1.0 $\mu$m and supercurrents have been measured over unprecedented lengths of 6 $\mu$m. d | In superconducting quantum point contacts, discretization of switching current $I_S$ (top panel) and subgap conductance $G_S$ (bottom panel) have been reported, demonstrating ballistic transport in superconducting devices. e | In superconducting quantum interference devices, periodic oscillations in the switching current are observed as a function of the out-of plane magnetic field. Panel a is reproduced with permission from ref~\cite{xiang_gesi_2006}, Springer Nature Limited. Panel b is reproduced with permission from ref~\cite{sistani_highly_2019}, American Chemical Society. Panel c and d are reproduced with permission from ref~\cite{hendrickx_ballistic_2019,hendrickx_gate-controlled_2018}, Springer Nature Limited. Panel e is reproduced with permission from ref~\cite{vigneau_germanium_2019}, American Chemical Society.}
\label{hybrids}
\end{figure}

The first experimental demonstration of a JoFET in Ge was based on a single Ge/Si core/shell NW connecting two superconducting aluminium electrodes \cite{xiang_gesi_2006} (Fig.~\ref{hybrids}a).  Well below 1 K, i.e. well below the superconducting critical temperature of the Al contacts, the device exhibited a superconducting regime with a gate-dependent critical current and an $I_c R_N$ product close to the ideal value, denoting a very high transparency of the contacts ($R_N$ is the normal-state resistance). The experimental demonstration of high-quality proximitized Ge/Si NWs and the recent measurements of a strong direct Rashba spin-orbit coupling\cite{Vries2018,Sun2018} make 1DHG Ge/Si core/shell NWs a favourable system for the realization of topological superconductivity and MZM\cite{Kloeffel2011,Maier2014b}. Furthermore, novel S-Sm devices structures such as a linear array of superconducting quantum dots were proposed to realize robust and practical Majorana chains\cite{sau_realizing_2012}. Ge-based superconducting quantum dots were pioneered in strained SiGe self-assembled nanocrystals\cite{katsaros_hybrid_2010}. The recent demonstration of superconducting quantum dots in Ge/Si core/shell NWs\cite{ridderbosAdvMat2018,Vries2018} represents an important step towards these alternative architectures for MZM.

The observation of a tunable superconducting proximity effect in Ge/Si core/shell NWs was reproduced in more recent experiments using not only Al \cite{Vries2018, ridderbos_multiple_2019, sistani_highly_2019} but also NbTiN \cite{kotekar-patil_quasiballistic_2017} superconducting contacts.  The high transparency of the S-Sm contacts was further confirmed by tunnel spectroscopy measurements \cite{ridderbos_hard_2020}, which revealed an induced superconducting gap containing only a small density of quasiparticle states. This experimental finding is particularly important in the prospect of creating Majorana edge states whose potentially long lifetime is not ruined by quasiparticle poisoning. 

Besides Ge/Si NWs, JoFETs in Ge were recently demonstrated in accumulation-only undoped Ge/SiGe heterostructrutures by contacting the induced two-dimensional hole gas with aluminum electrodes\cite{hendrickx_gate-controlled_2018,hendrickx_ballistic_2019, vigneau_germanium_2019}. Owing to the rather high hole mobility, supercurrent transport was observed even for Ge channels as long as 6 $\mu$m\cite{ hendrickx_ballistic_2019} (Fig.~\ref{hybrids}c). Compared to Ge/Si NWs, planar Ge/SiGe heterostructures are a more versatile material platform offering ample freedom for device design. Figure ~\ref{hybrids}d shows the scanning electron micrograph of an Al-Ge-Al junction embedding an additional pair of split gates defining a quantum point contact. Upon varying the voltage on these gates, the switching current and sub-gap conductance exhibited a step-like structure reflecting the discretized opening of one-dimensional hole modes\cite{hendrickx_ballistic_2019}. Figure ~\ref{hybrids}e reproduces another example of hybrid S-Sm device consisting of two independently controlled JoFETs fabricated on the same mesa structure and embedded in an Al superconducting ring \cite{vigneau_germanium_2019}. This geometry realizes a gate-tunable superconducting quantum interference device.

We have already stressed the importance of having S-Sm junctions highly transparent to the flow of carriers between the semiconductor and the superconductor.  To this aim, the possible presence of insulating interface layers, such as native oxides, needs to be avoided. The \textit{in-situ} deposition of the superconducting metal, right after the semiconductor growth, offers an effective solution \cite{de_franceschi_andreev_1998, krogstrup_epitaxy_2015}. Interestingly, Ge-based structures allow also for an equally valuable \textit{ex-situ} approach relying on the thermally activated propagation of the superconducting element into the Ge channel. This process was clearly demonstrated for Ge-based nanowires contacted by Al electrodes \cite{kral_abrupt_2015}. Transmission tunnelling microscopy carried out during the annealing process shows aluminium entering the nanowire and replacing Ge \cite{el_hajraoui_situ_2019}. This solid-state reaction results in a clean, atomically sharp Al/Ge interface progressively moving into the Ge nanowire. Because of its low activation energy, this process can occur for annealing temperatures as low as 250 $^\circ$C. Considering that device fabrication involves processing steps at comparably high temperatures (e.g. atomic-layer deposition), the inner diffusion of aluminium is probably at the origin of the high contact transparencies observed in most of the experiments discussed above. Recently, Al-Ge-Al devices with an extremely short Ge channel could be obtained by a controlled annealing process \cite{sistani_highly_2019} (Fig.~\ref{hybrids}b). The implementation of this contact fabrication method to other superconducting metals, having higher critical temperature and higher critical magnetic fields, would further enlarge the range of device functionalities accessible with Ge-based hybrid systems.

\section*{Outlook}

After two decades of research on quantum computation with quantum dots\cite{Loss1998}, the ingredients for extensible qubit tiles are becoming concrete\cite{vandersypen_interfacing_2017,franke_rents_2019} and several appealing architectures have been proposed\cite{taylor_fault-tolerant_2005,veldhorst_silicon_2017,li_crossbar_2018,hill_surface_2015}. Attractive architectures for large-scale quantum computing are based on qubit modules, consisting of linear or two-dimensional arrays, interconnected using long-range links\cite{vandersypen_interfacing_2017}. While one may expect that the interconnections themselves will form the bottleneck, quantum error correction schemes show that remarkable error rates can be tolerated when the qubit modules can execute quantum logic with high-fidelity\cite{nickerson_topological_2013}. Great progress has been made with quantum dot qubits: universal gates sets for quantum logic\cite{veldhorst_two-qubit_2015, hendrickx_fast_2020}, rudimentary quantum algorithms\cite{watson_programmable_2018}, electron spin states using gate-based readout \cite{zheng_rapid_2019,west_gate-based_2019,urdampilleta_gate-based_2019,crippa_gate-reflectometry_2019}, strong coupling between spin and microwave photons\cite{samkharadze_strong_2018,mi_coherent_2018}, and interactions between distant electron spins\cite{borjans_resonant_2020}. The grand challenge is to integrate these individual elements on a single platform to build scalable quantum technology. In this respect, Ge is particularly appealing since it may allow for the co-integration of different quantum components relying on spin, superconductivity, or, possibly, topologically protected quasiparticle states. Here we present an outline of these possible quantum components and identify some of these exciting pursuits that lay on the road for Ge quantum electronics based on nanowires, hut wires and planar systems.

\begin{figure}[ht!]
\centering
\includegraphics[width=110mm]{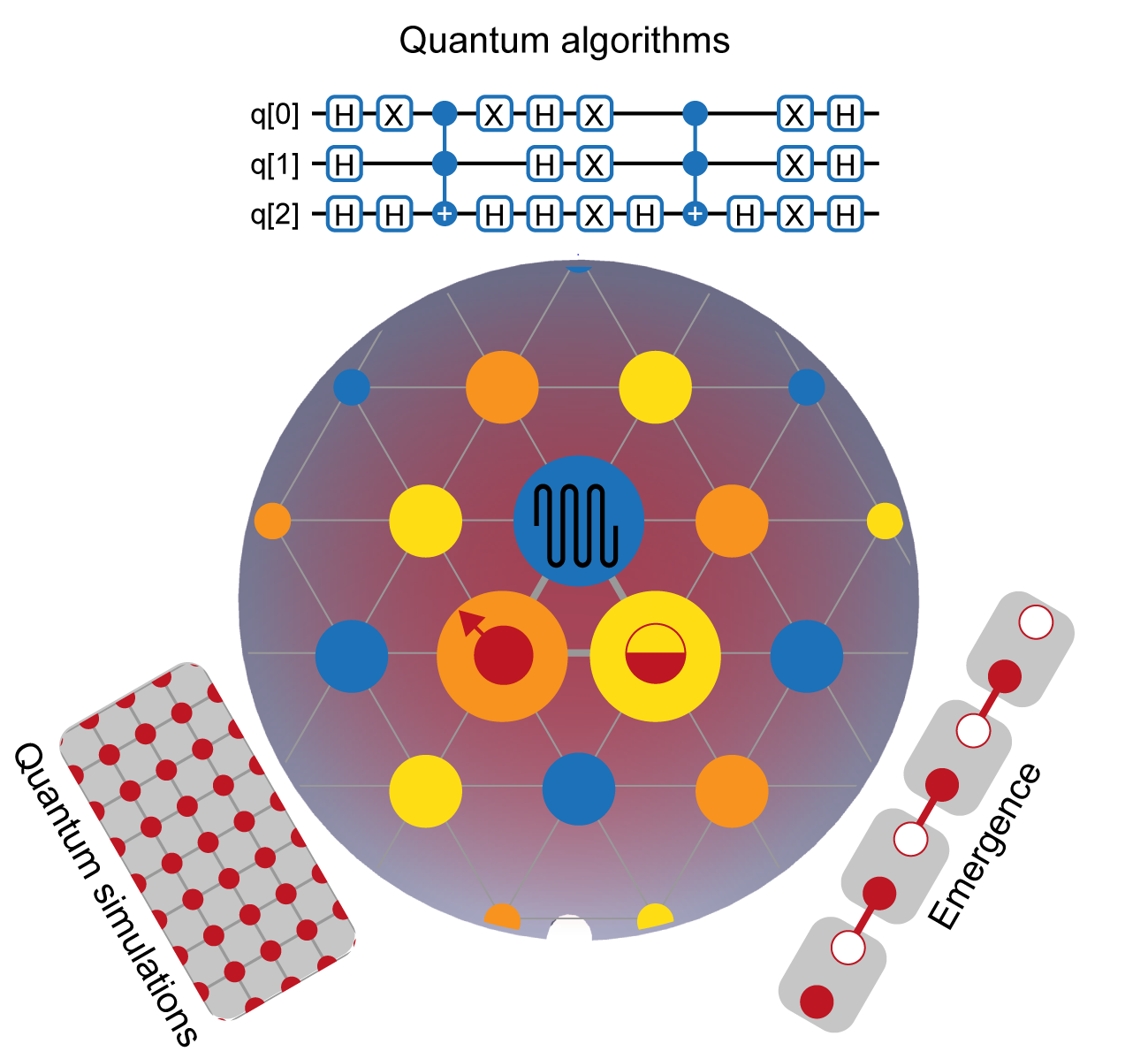}
\caption{\textbf{Germanium-based quantum technology} The extremely rich Ge platform supports the development of spin, superconductor, and topological systems. While each of these systems defines an exciting direction that may support high-fidelity qubits and quantum logic, their co-integration in Ge provides an opportunity for scalable quantum technology on a Si wafer. Hybrid quantum systems offer a unique platform for the simulation of non-trivial hamiltonians, such as the Fermi-Hubbard system, and allow for the controllable emulation of emergent physics such as Majorana zero modes in $p$-wave superconductors. For fault-tolerant quantum computation, topological regions may function as error-protected long-range links for spin qubits, while single qubit rotations on gatemons or spin qubits can complement operations on topological qubits to define a universal quantum gate set.}
\label{perspective}
\end{figure}

\subsection*{Logical qubits based on all-electrically driven spin qubits}
The all-electric control of hole spins mediated by the spin-orbit coupling has already led to the operation of a universal gate set, by combining single \cite{Watzinger2018, hendrickx_fast_2020} and two-qubit logic \cite{hendrickx_fast_2020}. It furthermore provides a pathway to scale hole quantum dot systems in two-dimensions required for quantum error correction\cite{fowler_surface_2012}. As opposed to electron spin qubits, which require large objects such as striplines or magnets to coherently control the electron spin \cite{vandersypen_interfacing_2017}, hole quantum dots have all elements for operation intrinsically available. This can largely facilitate scalability in two-dimensions. Indeed, first demonstrations of 2D quantum dot arrays in Ge have already been reported \cite{lawrie_quantum_2020}. Furthermore, Ge can be purified to contain only isotopes with zero nuclear spins to remove nuclear spin batch dephasing and allow for long quantum coherence times\cite{Itoh1993,itoh_isotope_2014}. This provides great promise to engineer quantum dot systems where quantum error correction schemes can be executed to build logical qubits with an error rate below the physical qubit.

\subsection*{Fast and high-fidelity gate-based readout of single hole spins}
The strong spin-orbit coupling for holes provides also a natural advantage for gate-based readout of single spins. Research from several groups have shown the promise of this approach using electron spins in Si. \cite{zheng_rapid_2019,west_gate-based_2019,urdampilleta_gate-based_2019} Adapting this technology to holes in Ge may further increase the readout rate as well as the readout fidelity. Indeed first steps have already been taken by showing fast rf readout\cite{lawrie_quantum_2020, hofmann2019assessing} and capacitive coupling\cite{Xu2019}.

\subsection*{Long-distance entanglement between single hole spins}
Single hole spins may be coherently coupled to microwave photons using on-chip superconducting resonators. The presence of strong spin orbit coupling can act as an efficient mediator to achieve spin-photon coupling\cite{hu_strong_2012} with rates beyond those achieved with electrons \cite{samkharadze_strong_2018,mi_coherent_2018} and first steps toward such coupling seem encouraging\cite{Xu2019}. By coupling two hole spins to a single resonator and bringing them into resonance, one can entangle hole qubits separated far apart. Such a realization would mean a milestone toward the realization of extensible qubit tiles coupled via long-range links. 

\subsection*{Highly tunable and low-noise superconducting gatemons}
Holes in Ge provide more surprises, since many metals have a Fermi level pinning to the valence band of Ge \cite{Dimoulas2006}, meaning that Schottky-barrier-free contacts can be made to superconductors. Gate-tunable superconductivity has been reported for nanowires \cite{xiangnnano2006} and planar systems\cite{hendrickx_gate-controlled_2018} and SQUIDs have already been realized \cite{vigneau_germanium_2019}. This opens exciting perspectives to build gatemons\cite{larsen_semiconductor-nanowire-based_2015,de_lange_realization_2015}. While gatemons may form the basis of a scalable qubit tile, in particular when based on two-dimensional systems\cite{casparis_superconducting_2018}, they could also function as a readout mechanism and as long-range link to couple distant hole spin qubits.

\subsection*{Topological qubits based on Majorana zero modes}
Unconventional superconductivity has been studied for a long time\cite{sigrist_phenomenological_1991}, but recently gained even more attention after the prediction that exotic superconductivity may emerge in superconductor-semiconductor systems\cite{sau_generic_2010,alicea_majorana_2010}. In particular, the combination of strong spin-orbit coupling, magnetic field, and superconductivity may lead to superconductivity with $p$-wave symmetry, providing an experimental testbed to emulate the original Kitaev chain\cite{Kitaev_2001}. In contrast to the common BCS $s$-wave superconductors, superconductors with $p$-wave symmetry may host zero energy states that can be described as Majorana fermions, and these states could exhibit non-Abelian exchange statistics\cite{read_paired_2000} and form the basis of topological quantum computation\cite{nayak_non-abelian_2008}. While several highly exciting experiments have been performed using InAs and InSb nano structures\cite{lutchyn_majorana_2018}, definite proof of the existence of isolated MZM has yet to be demonstrated. Holes in Ge can exhibit strong spin-orbit coupling and Ge quantum devices may therefore form an excellent host for MZM\cite{Maier2014,Maier2014b}. Furthermore, by introducing a novel Berry phase\cite{mao_superconducting_2011}, heavy holes may provide a new twist to Majorana states.

\subsection*{Quantum information transfer between different qubit types}
Germanium turns out to be an exciting material to explore several qubit types. It also provides the opportunity to study the interaction between different qubits. Perhaps the most exciting one is the interaction between topologically trivial and topological qubits. Quantum information transfer may occur between spin qubits based on individual holes and parity qubits based on MZM\cite{leijnse_quantum_2011,leijnse_hybrid_2012,hoffman_universal_2016,Rancic2019}. While scientifically highly interesting, such a transfer may also resolve key challenges in quantum information. First, it could provide a topological link between separated hole qubits. Secondly, it may be used to construct a universal gate set for topological qubits. Topological qubits based on MZM can only be used to construct states within the Clifford group, which can be efficiently simulated on a classical computer\cite{nayak_non-abelian_2008}. The quantum gate set may be enriched by exploiting the spin qubit toolbox to enable full universal logic. While the realization would represent a formidable achievement in topological quantum computing, significant steps have to be taken. Nonetheless, exciting intermediate steps can be expected along the way. These include the coupling of superconductors to spin qubits to enable superconductivity-mediated long-range coupling between spin qubits based on crossed Andreev reflection\cite{choi_spin-dependent_2000,leijnse_coupling_2013}, with interactions predicted to exceed micrometers in one-dimensional systems\cite{hassler_exchange_2015}.

\bibliography{sample}

\noindent\textbf{Acknowledgements}
G.S., M.W.,F.A.Z acknowledge financial support from The Netherlands Organization for Scientific Research (NWO). F.Z., D.L., G.K. acknowledge funding from the European Union's Horizon 2020 research and innovation programme under Grand Agreement Nr. 862046. G.K. acknowledges funding from FP7 ERC Starting Grant 335497, FWF Y 715-N30, FWF P-30207. S.D. acknowledges support from the European Union’s Horizon 2020 program under Grant Agreement No. 81050 and from the Agence Nationale de la Recherche through the TOPONANO and CMOSQSPIN projects. J.Z. acknowledges support from the National Key R\&D Program of China (Grant No. 2016YFA0301701) and Strategic Priority Research Program of CAS (Grant No. XDB30000000). D.L. and C.K. acknowledge the Swiss National Science Foundation and NCCR QSIT.

\noindent\textbf{Author contributions}\\
All authors contributed to the writing of the manuscript.

\noindent\textbf{Competing interests}\\
The authors declare no competing interests.

\end{document}